\documentclass[aps,prl,floatfix,twocolumn,nofootinbib,superscriptaddress,longbibliography]{revtex4-2}

\usepackage{graphicx}

\usepackage{todonotes}
\usepackage{float}
\usepackage{orcidlink}
\usepackage[version=4]{mhchem}
\usepackage{dcolumn}
\usepackage{bm}
\usepackage{hyperref}
\hypersetup{
   breaklinks=true,   
}
\begin{document}
\preprint{APS/123-QED}
\title{The mass of $^{101}$Sn and Bayesian extrapolations to the proton drip line}%

\author{C.~M.~Ireland\,\orcidlink{0009-0002-4079-1567}}
\affiliation{Facility for Rare Isotope Beams, Michigan State University, East Lansing, Michigan 48824, USA}
\affiliation{Department of Physics and Astronomy, Michigan State University, East Lansing, Michigan 48824, USA}
\email{ireland@frib.msu.edu}

\author{G.~Bollen\,\orcidlink{0000-0003-4453-1180}}
\affiliation{Facility for Rare Isotope Beams, Michigan State University, East Lansing, Michigan 48824, USA}
\affiliation{Department of Physics and Astronomy, Michigan State University, East Lansing, Michigan 48824, USA}

\author{S.~E.~Campbell\,\orcidlink{0009-0005-2053-2711}}
\affiliation{Facility for Rare Isotope Beams, Michigan State University, East Lansing, Michigan 48824, USA}
\affiliation{Department of Physics and Astronomy, Michigan State University, East Lansing, Michigan 48824, USA}

\author{X.~Chen\,\orcidlink{0000-0003-3513-8870}}
\affiliation{Facility for Rare Isotope Beams, Michigan State University, East Lansing, Michigan 48824, USA}

\author{H.~Erington\,\orcidlink{0009-0005-4433-3020}}
\affiliation{Facility for Rare Isotope Beams, Michigan State University, East Lansing, Michigan 48824, USA}
\affiliation{Department of Physics and Astronomy, Michigan State University, East Lansing, Michigan 48824, USA}

\author{N.~D.~Gamage\,\orcidlink{0000-0001-5657-2081}}
\affiliation{Facility for Rare Isotope Beams, Michigan State University, East Lansing, Michigan 48824, USA}

\author{K.~Godbey\,\orcidlink{0000-0003-0622-3646}}
\affiliation{Facility for Rare Isotope Beams, Michigan State University, East Lansing, Michigan 48824, USA}

\author{A.~M.~Houff}
\affiliation{Department of Physics and Astronomy, University of Notre Dame, Notre Dame, Indiana, 46556, USA}

\author{C.~Izzo\,\orcidlink{0000-0002-4119-1654}}
\affiliation{Facility for Rare Isotope Beams, Michigan State University, East Lansing, Michigan 48824, USA}

\author{B.~Knight\,\orcidlink{0009-0004-6777-1034}}
\affiliation{Facility for Rare Isotope Beams, Michigan State University, East Lansing, Michigan 48824, USA}
\affiliation{Department of Physics and Astronomy, Michigan State University, East Lansing, Michigan 48824, USA}

\author{S.~Lalit\,\orcidlink{0000-0001-7758-492X}}
\affiliation{Facility for Rare Isotope Beams, Michigan State University, East Lansing, Michigan 48824, USA}

\author{E.~Leistenschneider\,\orcidlink{0000-0003-0377-8523}}
\affiliation{Nuclear Science Division, Lawrence Berkeley National Laboratory, Berkeley, California 94720, USA}

\author{E.~M.~Lykiardopoulou}
\affiliation{Nuclear Science Division, Lawrence Berkeley National Laboratory, Berkeley, California 94720, USA}

\author{F.~M.~Maier\,\orcidlink{0000-0001-9715-2147}}
\affiliation{Facility for Rare Isotope Beams, Michigan State University, East Lansing, Michigan 48824, USA}

\author{W.~Nazarewicz\,\orcidlink{0000-0002-8084-7425}}
\affiliation{Facility for Rare Isotope Beams, Michigan State University, East Lansing, Michigan 48824, USA}
\affiliation{Department of Physics and Astronomy, Michigan State University, East Lansing, Michigan 48824, USA}

\author{R.~Orford}
\affiliation{Nuclear Science Division, Lawrence Berkeley National Laboratory, Berkeley, California 94720, USA}

\author{W.~S.~Porter\,\orcidlink{0000-0002-5172-3298}}
\affiliation{Department of Physics and Astronomy, University of Notre Dame, Notre Dame, Indiana, 46556, USA}

\author{C.~Quick}
\affiliation{Department of Physics and Astronomy, University of Notre Dame, Notre Dame, Indiana, 46556, USA}

\author{A.~Ravli\'c\,\orcidlink{0000-0001-9639-5382}}
\affiliation{Facility for Rare Isotope Beams, Michigan State University, East Lansing, Michigan 48824, USA}
\affiliation{Department of Physics, Faculty of Science, University of Zagreb, Bijeni\v cka c. 32, 10000 Zagreb, Croatia}

\author{M.~Redshaw\,\orcidlink{0000-0002-3013-0100}}
\affiliation{Facility for Rare Isotope Beams, Michigan State University, East Lansing, Michigan 48824, USA}
\affiliation{Department of Physics, Central Michigan University, Mount Pleasant, Michigan 48859, USA}

\author{P.-G.~Reinhard\,\orcidlink{0000-0002-4505-1552}}
\affiliation{Institut für Theoretische Physik, Universität Erlangen, Erlangen D-91054, Germany}

\author{R.~Ringle\,\orcidlink{0000-0002-7478-259X}}
\affiliation{Facility for Rare Isotope Beams, Michigan State University, East Lansing, Michigan 48824, USA}
\affiliation{Department of Physics and Astronomy, Michigan State University, East Lansing, Michigan 48824, USA}

\author{S.~Schwarz\,\orcidlink{0000-0002-1522-5929}}
\affiliation{Facility for Rare Isotope Beams, Michigan State University, East Lansing, Michigan 48824, USA}

\author{C.~S.~Sumithrarachchi\,\orcidlink{0009-0009-3123-0937}}
\affiliation{Facility for Rare Isotope Beams, Michigan State University, East Lansing, Michigan 48824, USA}

\author{A.~A.~Valverde\,\orcidlink{0000-0002-9118-6589}}
\affiliation{Physics Division, Argonne National Laboratory, Lemont, Illinois, 60439, USA}

\author{A.~C.~C.~Villari\,\orcidlink{0000-0003-2914-7471}}
\affiliation{Facility for Rare Isotope Beams, Michigan State University, East Lansing, Michigan 48824, USA}

\date{\today}%

\begin{abstract}
The favorable energy configurations of nuclei at magic numbers of ${N}$ neutrons and ${Z}$ protons are fundamental for understanding the evolution of nuclear structure. The ${Z=50}$ (tin) isotopic chain is a frontier for such studies, with particular interest at and around the doubly-magic \textsuperscript{100}Sn isotope, for which the mass is a topic of debate. Precise mass values for neutron-deficient isotopes provide necessary anchor points for mass models to test extrapolations near the proton drip line, where experimental studies remain out of reach. In this work, we report the first Penning trap mass measurement of \textsuperscript{101}Sn. The determined mass excess of $-59\,889.89(96)$~keV for \textsuperscript{101}Sn represents a factor of 300 improvement over the current precision and indicates that \textsuperscript{101}Sn is less bound than previously thought. Mass predictions from a recently developed Bayesian model combination (BMC) framework employing statistical machine learning and nuclear masses computed within seven global models based on nuclear Density Functional Theory (DFT) agree within 1$\sigma$ with experimental masses from the $48 \le Z \le 52$ isotopic chains. The framework's resilience to new mass data gave confidence in the extrapolation of tin masses down to $N=46$. Our calculations suggest that \textsuperscript{96}Sn is a two-proton drip line nucleus and predict a mass excess of $-58\,090(800)$~keV for $^{100}$Sn, showing a preference within 1$\sigma$ for the mass of \textsuperscript{100}Sn derived from the $\beta$-delayed $Q$-value measured at GSI.
\end{abstract}

\maketitle

\textit{Introduction.}--Studying nuclei at magic numbers has been key to the development of the nuclear shell model~\cite{Otsuka2020,Dobaczewski2007}. As such, the properties of the tin isotopic chain near the doubly magic shell closure at \textsuperscript{100}Sn continue to generate interest~\cite{FAESTERMANN201385,physics4010024,Vernon2025,100Sn_chargeradii, Togashi_2025, wbdx-k3cd}. The studies of shell closures near the drip lines, isospin symmetry breaking, and proton-neutron pairing make the binding energy of \textsuperscript{100}Sn a crucial observable. Unfortunately, the low  production rates in this region hinder experimental studies. The initial mass measurement of \textsuperscript{100}Sn 
was prone to large systematic and statistical errors ($>1$~MeV)~\cite{Chartier1996}, and a near 2$\sigma$ discrepancy exists between the two reported $\beta$-decay $Q$-values for $^{100}$Sn from RIKEN~\cite{Lubos2019} and GSI~\cite{Hinke2012}. The dominant mass value reported in the 2020 Atomic Mass Evaluation (AME2020)~\cite{AME2020} is indirectly calculated from Ref.~\cite{Hinke2012}. 

Nuclei surrounding \textsuperscript{100}Sn are also of interest for both nuclear structure and astrophysics studies~\cite{Mougeot2021,PhysRevC.102.014304,Reponen2021,HORNUNG2020135200, Karthein2024, AURANEN2019187,PhysRevLett.102.252501}. Recently, mass measurements of $^{103}$Sn indirectly constrained the binding energies of several nuclei in this region~\cite{Ireland2025, Nies2025, StorageRing} and suggested a $300$~keV shift to the reported mass excess of \textsuperscript{101}Sn in AME2020~\cite{AME2020}. This would make $^{101}$Sn less bound than previously determined by the available $\beta$-delayed proton data~\cite{straub2010decay}. A precision mass measurement of $^{101}$Sn would test this suggestion, constrain the masses of the \textsuperscript{109}Xe$\rightarrow$\textsuperscript{105}Te$\rightarrow$\textsuperscript{101}Sn $\alpha$-decay chain, and further illuminate the 2$\sigma$ tension in the mass of $^{100}$Sn pointed out in Ref.~\cite{Mougeot2021}.

One can also leverage new measurements to extrapolate to unknown masses. The AME2020 relies on model-independent systematic trends and empirical relationships to make extrapolations encompassing only a few nuclei in the vicinity of the measured data~\cite{AME2020}.
To reliably extend further, predictive model-based approaches are required. Nuclei around $^{100}$Sn have been studied via the shell model~\cite{Grawe1995,Grawe2002,Darby2010,FAESTERMANN201385}, nuclear density functional theory (DFT)~\cite{Redon1988,DobNaz95}, and \textit{ab-initio} models~\cite{PhysRevLett.120.152503, PhysRevC.104.064310,Mougeot2021, hildenbrand2025lattice}. Considering that all nuclear mass models are imperfect, one cannot rely on a single model when extrapolating into an unknown domain of $(Z,N)$. Instead, one can adopt a principled Bayesian model mixing framework informed by current experimental data. Such a multi-model inference has been used to generate quantified mass extrapolations~\cite{Neufcourt2018,Neufcourt2019,Neufcourt2020,Neufcourt2020b,Kejzlar2023,Saito2024,Giuliani2024,Wu2024} that placed
the $Z=50$ two-proton drip line at \textsuperscript{97}Sn~\cite{Neufcourt2020}.
Crucial to these approaches is that `prediction' is defined by a probability that can be updated with new experimental data, without being overly sensitive and prone to overfitting.
In this regard, the recently developed Bayesian Model Combination (BMC) framework~\cite{Giuliani2024} has emerged as a powerful data science technique for continuous integration of precision measurements. Given that for any extrapolation with a trustworthy credible interval, comparing a single prediction or the mean result of a quantified model masks the underlying variability arising from errors in the data/model, the predictive power can be ascertained only by verifying the resilience of extrapolations to new data. Therefore, a study around $^{101}$Sn represents the first application of the BMC framework to real-world mass data of rare-isotopes. Additionally, there is interest pertaining to the ground state assignment of \textsuperscript{101}Sn~\cite{seweryniak2007single, Darby2010, yakhelef2024neutron,PhysRevC.102.014304}. While a precise mass measurement of \textsuperscript{101}Sn will not fully resolve this, state-of-the-art theoretical predictions aided by this measurement could provide insight into this discussion and suggestions of the level ordering.

\textit{Experimental procedure and analysis.}--Neutron-deficient \textsuperscript{101}Sn was produced at the Facility for Rare Isotope Beams (FRIB) via projectile fragmentation. A $^{124}$Xe primary beam was accelerated to an energy of 228 MeV/u using the superconducting linear accelerator~\cite{York-LINAC} and impinged upon a $\approx$2~mm thick $^{12}$C target. The resulting isotope cocktail was sent through the Advanced Rare Isotope Separator (ARIS)~\cite{PORTILLO2023151} which partially purified and identified the beam components. The purified beam then passed through a momentum compression beamline to achieve efficient stopping in the Advanced Cryogenic Gas Stopper (ACGS)~\cite{Lund-ACGS} where ions were stopped via helium buffer gas collisions, extracted, and reaccelerated. Following mass filtering through a dipole magnet with a resolving power of $\approx$1500, ions with a mass-to-charge ratio $A/$q=50.5, including the doubly-charged \textsuperscript{101}Sn\textsuperscript{2+}, were selected and sent to LEBIT with an energy of 30~keV$\cdot$q. The continuous ion beam was then injected into the buffer-gas-filled cooler-buncher~\cite{CoolerBuncher}, where the ions were cooled, bunched, and extracted. Ion bunches were then processed by a coarse time-of-flight filter before being transferred to LEBIT’s 9.4~T Penning trap mass spectrometer~\cite{PenningTrap}.
\newline \indent The mass~$m$ of an ion with charge~$q$ stored in a Penning trap with a magnetic field $B$ can be determined by measuring its cyclotron frequency $\nu_{c}=qB/(2\pi m) = \nu_{+} + \nu_{-}$~\cite{Gabrielse-Sideband}. Radial frequency measurements of the reduced cyclotron ($\nu_{+}$) and magnetron ($\nu_{-}$) frequencies were performed using the phase-imaging ion-cyclotron-resonance (PI-ICR) technique~\cite{PIICR1, PIICR2}, following the approach outlined in Ref.~\cite{PIICR2}, which independently measures the radial frequencies via phase advances on a downstream position-sensitive detector after excited ions evolve for an accumulation time $t_{\mathrm{acc}}$ in the trap and are extracted. The relevant radial frequency is given by
\begin{equation}\label{eq:pi}
    \nu_{\pm} = \frac{\left(\phi_{\mathrm{acc}_{\pm}} - \phi_{\mathrm{ref}}\right) + 2\pi n_{\pm}}{2\pi t_{\mathrm{acc}_{\pm}}},
\end{equation}
where $t_{\mathrm{acc}_{\pm}}$ and $\phi_{\mathrm{acc}_{\pm}}$ refer to the accumulation time and phase advance relative to a reference location $\phi_{\mathrm{ref}}$ for either $\nu_{+}$ or $\nu_{-}$ motion, depending on the excitation scheme described above. From Eq.~\ref{eq:pi}, it is apparent that longer accumulation times result in higher precision frequency measurements. Assuming an initial frequency uncertainty of $\delta_{\nu_{\pm}}<1/t_{\mathrm{acc}_{\pm}}$ for a prior reported frequency $\nu_{\pm,\mathrm{initial}}$, the number of complete revolutions undergone in the trap $n_{\pm}$ is known exactly by $\lfloor\nu_{\pm,\mathrm{initial}} t_{\mathrm{acc}_{\pm}}\rfloor$, with the remaining partial revolution determining the precise frequency. 

\begin{figure}
        \includegraphics[width=1\linewidth]{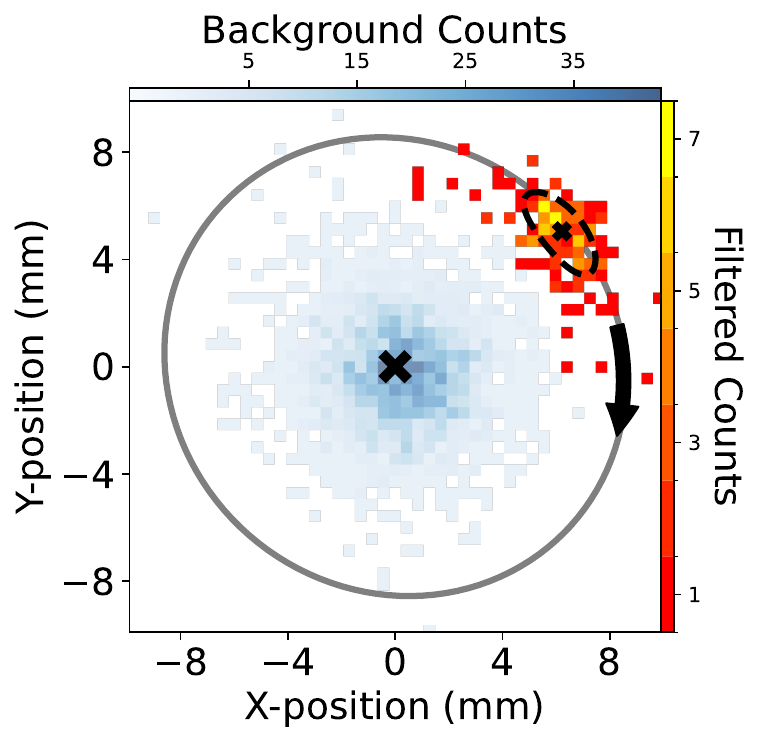}
        \caption{A filtered visual of $\approx$150 detected \textsuperscript{101}Sn\textsuperscript{2+} ions combined from three separate frequency measurements after ions accumulated phase while undergoing 163~ms of reduced-cyclotron motion is highlighted alongside unidentified background counts. The direction of motion along the gray elliptical path is shown with a black arrow and the fitted center for filtered counts is shown with a large black \textbf{x}. The filtered ions are fit to a two-dimensional Gaussian distribution in polar coordinates, where the center of the fit and $1\sigma$ error boundary are indicated by a small black \textbf{x} and a dashed black ellipse respectively.}
        \label{fig:PI}
\end{figure} 

Identifying the trapped ions requires calibration of the magnetic field. To do so, measurements of \textsuperscript{101}Sn\textsuperscript{2+} were interwoven with cyclotron frequency measurements of \textsuperscript{50}Ti\textsuperscript{+} and \textsuperscript{51}V\textsuperscript{+} produced by LEBIT's offline laser ion source~\cite{izzo2016laser}. The well-known mass ratio of $^{51}$V to $^{50}$Ti provides the desired relationship between the cyclotron frequency and the mass of trapped ions. The large prior mass uncertainty for \textsuperscript{101}Sn\textsuperscript{2+} required that initial measurements of $\nu_{+}$ used lower accumulation times of $t_{\mathrm{acc}_{+}}=$~5.5 and 13.8~ms to differentiate the expected angular location of \textsuperscript{101}Sn\textsuperscript{2+} from that of any radioactive/stable isobars that could have been present. The obtained frequencies were cross-checked against masses in the AME2020 database, including any molecular ions. No candidates except for \textsuperscript{101}Sn\textsuperscript{2+} were plausible, providing a positive identification. The resulting frequency uncertainty of $\delta_{\nu_{+}}\approx1$Hz instilled confidence in $n_{+}$ for higher accumulation times, allowing measurements with $t_{\mathrm{acc}_{+}}=$~50, 150, and 163~ms to reduce the measured mass uncertainty. An example of detected counts of \textsuperscript{101}Sn\textsuperscript{2+} alongside unidentified contaminants is shown in Figure~\ref{fig:PI}. Due to the weak mass dependence of the magnetron frequency~\cite{Gabrielse-Sideband}, the $\nu_{-}$ of \textsuperscript{101}Sn\textsuperscript{2+} was linearly interpolated from that of \textsuperscript{50}Ti\textsuperscript{+} and \textsuperscript{51}V\textsuperscript{+}. Individual recorded clusters of \textsuperscript{101}Sn\textsuperscript{2+} accumulated 30-50~counts while calibrant species clusters accumulated $\approx$200 counts. Calibrant measurements utilized phase accumulation times of $t_{\mathrm{acc}_{-}}$~=~181~ms and $t_{\mathrm{acc}_{+}}$~=~197~ms. The mass was then extracted from the cyclotron frequency ratio $R$ of \textsuperscript{101}Sn\textsuperscript{2+} relative to \textsuperscript{50}Ti\textsuperscript{+},
\begin{equation}\label{eq:mass_from_ratio}
   R = \frac{ \nu_{c} }{ \nu_{c,\mathrm{calib}} } = \frac{ q  m_{\mathrm{calib}} }{ q_{\mathrm{calib}} m }  ,
\end{equation}
where the charges $q$ and $q_{\mathrm{calib}}$ and the masses $m$ and $m_{\mathrm{calib}}$ refer to \textsuperscript{101}Sn\textsuperscript{2+} and \textsuperscript{50}Ti\textsuperscript{+}, respectively. \textsuperscript{50}Ti\textsuperscript{+} was chosen due to its better-known mass precision~\cite{AME2020}. Repeated, interwoven measurements of \textsuperscript{101}Sn\textsuperscript{2+} and both calibrants yielded a weighted average ratio $\bar{R}$.

Systematic effects associated with Penning trap mass spectrometry add shifts $\delta_{R}$ to~$\bar{R}$. As the detected rate of \textsuperscript{101}Sn\textsuperscript{2+} was $\approx$1 ion every 5 minutes, a complete mass measurement comprising both calibrant and ion of interest measurements lasted 7-8~hours. Non-linear magnetic field instabilities on these time scales~\cite{MagneticFieldShift} were negligible compared to the statistical uncertainty of $\delta_{R}\gtrsim 10^{-8}$. Shifts from isobaric contaminants~\cite{Bollen-IsomerRes} and ion-ion interactions~\cite{ToF1} for calibrant measurements were mitigated by short trapping times and keeping the number of detected ions per shot to $<$5. Linear mass-dependent shifts were treated with regular calibrant measurements. The systematic shift from the half-mass difference of \textsuperscript{101}Sn\textsuperscript{2+} and \textsuperscript{50}Ti\textsuperscript{+} was determined to be $\delta_{R}=-2(2)\times10^{-9}$ post-experiment, in line with the study in Ref.~\cite{PenningTrap}. This was added in quadrature to the statistical uncertainty. No shift in the mass of \textsuperscript{101}Sn was found when \textsuperscript{51}V\textsuperscript{+} was used as the calibrant in Eq.~\ref{eq:mass_from_ratio}.

Systematic effects related to PI-ICR have been studied in detail~\cite{Orford-CPTSys, Karthein-PIAnalysis, Nesterenko-JYFLPISys, Chenmarev-TRIGASys}. Specifically, residual magnetron motion due to improper ion injection onto the central axis of the trap was mitigated using a Lorentz steerer~\cite{LorentzSteerer} and averaging over the expected magnetron period of the ions prior to excitation. Trapping field misalignments additionally yield distorted projections of ion motion. Thus, the detected center of motion relative to the trap center was determined by periodically fitting an ellipse to the projected magnetron path of the ions on the detector (see Fig.~\ref{fig:PI}). No drift in the fitted center or the reference location of excited ions was observed. The optimal angle for placing ions onto the detector was determined to be $+45$~degrees by offline measurements that minimized the angular standard deviation of detected ions. After initial $\nu_{+}$ measurements with clusters located around 320~degrees confirmed the presence of \textsuperscript{101}Sn\textsuperscript{2+}, certainty in $n_{+}$ allowed for precise positioning near $+45$~degrees. Studies of \textsuperscript{50}Ti\textsuperscript{+} post-measurement showed no impact on the reported mass values from the measurements performed at 320~degrees.

\textit{Results.}--Eight PI-ICR measurements of \textsuperscript{101}Sn were taken over the course of 60 hours of experimental beam time. The well-known frequency ratio between \textsuperscript{50}Ti\textsuperscript{+} and \textsuperscript{51}V\textsuperscript{+} was periodically checked to confirm the system's stability. The weighted average ratio of these measurements $\bar{R}=0.989635524(10)$ corresponds to a mass excess $ME=-59\,889.89(96)$~keV for \textsuperscript{101}Sn, representing a factor of 300 improvement on the value reported in AME2020 of $-60\,310(300)$~keV~\cite{AME2020}. Individually measured frequency ratios are compared to $\bar{R}$ in Fig. S1 of the Supplementary Material~\cite{Sup}. This result is in agreement with the suggested adjustment from mass measurements of \textsuperscript{103}Sn~\cite{Ireland2025,Nies2025,StorageRing} and removes the remaining kink at $N=53$ in the two-neutron separation energy for the tin isotopic chain. Our achieved mass precision for \textsuperscript{101}Sn greatly reduces the mass uncertainties of the remaining nuclei in the \textsuperscript{109}Xe$\rightarrow$\textsuperscript{105}Te$\rightarrow$\textsuperscript{101}Sn $\alpha$-decay chain, derived from their $Q_{\alpha}$ values reported in Ref.~\cite{AME2020} and our measured mass for \textsuperscript{101}Sn. Table~\ref{tab: MassExcessTable} displays these results in comparison to those reported in AME2020.

\begin{table}[h!] 
\centering
\caption{Mass excess (in keV) for
$^{51}$V and $^{101}$Sn (using $^{50}$Ti as the calibrant), and $^{105}$Te
and $^{109}$Xe from the $\alpha$-decay of $^{101}$Sn from this work compared to values listed in AME2020~\cite{AME2020}.}
\vspace{\baselineskip}
\renewcommand{\arraystretch}{1.25}
\setlength{\tabcolsep}{12pt}
\begin{tabular}{c c c}
\hline
\hline
 Nucleus & $ME_{\mathrm{LEBIT}}$ & $ME_{\mathrm{AME2020}}$ \\ \hline
\textbf{$^{51}$V}    & $-52\,203.10(24)$ & $-52\,203.11(10)$ \\ 
\textbf{$^{101}$Sn}    & $-59\,889.89(96)$ & $-60\,310(300)$ \\ 
\textbf{$^{105}$Te}    &$-52\,396(3.2)$ & $-52\,810(300)$\\ 
\textbf{$^{109}$Xe}    &$-45\,754(8)$ & $-46\,170(300)$ \\ 
\hline
\hline
\end{tabular}
\label{tab: MassExcessTable}
\end{table}

Taking the mass excess for $^{101}$Sn and the average ground state mass of \textsuperscript{101}In reported in Ref.~\cite{Mougeot2021}, the electron-capture decay energy $Q_{EC}$ can be calculated using $Q_{EC}=M(^{101}\mathrm{Sn})-M(^{101}\mathrm{In})-B_e$, where $M$ is the atomic mass of the given nucleus and $B_e$ is the atomic binding energy of the $K$-shell electron in ${}^{101}$In~\cite{Bearden1967a}. The newly determined decay energy of $Q_{EC}=$~8626.1(13)~keV and the one-proton separation energy $S_{1p}$ of \textsuperscript{101}In from Ref.~\cite{Mougeot2021} allow us to deduce the available energy for $\beta$-delayed proton emission through \textsuperscript{101}In into \textsuperscript{100}Cd ($Q_{EC}-S_{1p}$) to be 6987(13)~keV, compared to 6600(300)~keV from Ref.~\cite{straub2010decay}, using the energy spectrum of $\beta$-delayed protons.

\begin{figure}[!t]
        \includegraphics[width=1.0\linewidth]{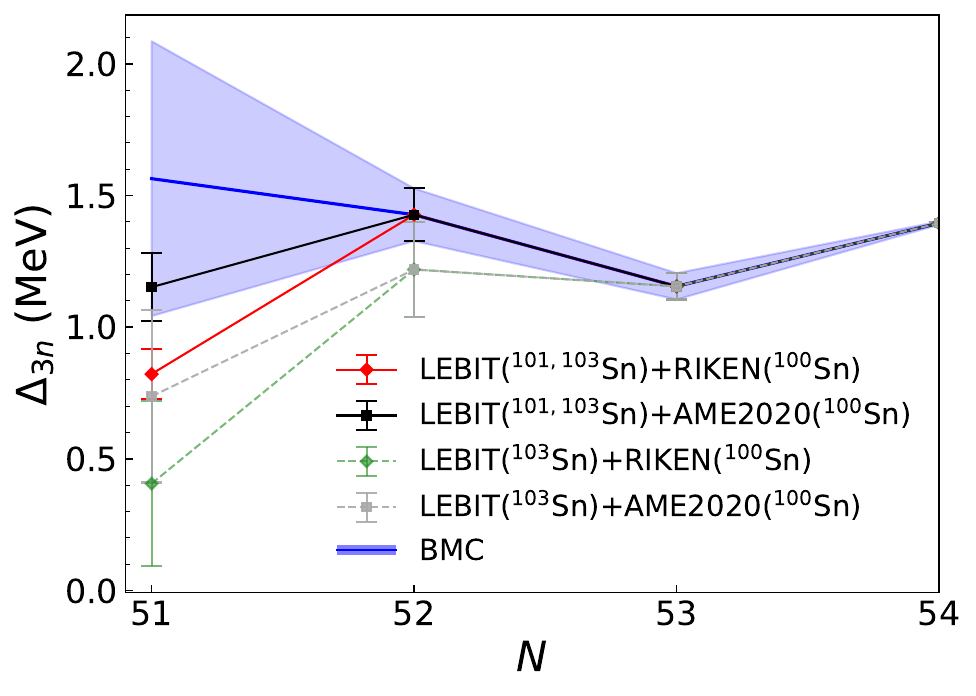}
        \caption{$\Delta_{3n}$ for the tin isotopic chain, displaying the discrepancy in the mass of $^{100}$Sn both before (dashed gray and green line) and after (solid black and red line) the mass measurement of $^{101}$Sn reported in this work via the odd-even staggering of the tin isotopic chain. For $N=51$, the mass of $^{100}$Sn is extracted using the mass of $^{100}$In from Ref.~\cite{Mougeot2021} and either the $\beta$-decay $Q$-value from Ref.~\cite{Hinke2012}, or Ref.~\cite{Lubos2019} (RIKEN). The ${}^{100}$Sn mass reported in AME2020\cite{AME2020} is based on the decay measurement from Ref. \cite{Hinke2012}.  All masses unspecified in the legend are taken from AME2020. The BMC predictions (solid blue) along with 1$\sigma$ uncertainty are also shown.}
        \label{fig:3pointLEBIT}
\end{figure} 

The mass of $^{101}$Sn impacts the three-point estimator for the neutron odd-even staggering $\Delta_{3n}(Z,N)= 0.5(-1)^N [ME(Z, N-1) - 2ME(Z,N) + ME(Z,N+1)]$ for $Z=50$, as shown in Fig.~\ref{fig:3pointLEBIT}. The tension in $\Delta_{3n}$, discussed in Ref.~\cite{Mougeot2021}, has been further substantiated by the improved mass precision for $^{101}$Sn. Indeed, the values of $\Delta_{3n}$
for $^{101}$Sn  based the $\beta$-delayed $Q$-values for $^{100}$Sn of Refs.~\cite{Lubos2019, Hinke2012} are now inconsistent (the  corresponding  $1\sigma$ errors do not overlap).

\textit{Bayesian mass extrapolations.}--With the new mass of $^{101}$Sn, we conduct a local Bayesian analysis around $^{100}$Sn using the BMC framework of Ref.~\cite{Giuliani2024} based on seven global energy density functionals (EDFs). In particular, the dataset includes three covariant EDFs: DD-PC1~\cite{Niksic2008}, DD-PCX~\cite{Yuksel2019} and DD-ME2~\cite{Lalazissis2005}, two Skyrme EDFs: UNEDF1~\cite{Kortelainen2012a} and UNEDF2~\cite{Kortelainen2014a}, as well as two Fayans EDFs: Fy($\Delta r$,HFB)~\cite{Reinhard2017a} and Fy(IVP)~\cite{Karthein2024}. These models meet the ``reasonable model'' condition of Ref.~\cite{Giuliani2024}. 
All calculations assume an axially-deformed, reflection-invariant nuclear shape. Calculations of odd-$A$ nuclei are performed through self-consistent blocking of the optimal quasiparticle configurations, on top of the reference vacuum of the even-even nucleus~\cite{Perez-Martin2008a}. Starting from the even-even vacuum, a set of one-quasiparticle candidate states in the vicinity of the Fermi level is identified for which the optimal shape of the nucleus was determined. Considering that the calculations are performed in the neighborhood of $^{100}$Sn, the Wigner energy term, using the prescription of Refs.~\cite{Buskirk2024,Hamaker2021a}, is employed to provide additional binding. One- and two-nucleon separation energies for $Z=50$ calculated using the individual EDFs are provided in Fig. S2 of the Supplementary Material~\cite{Sup}.

The BMC framework and its benefits over its constituent models are detailed in Ref.~\cite{Giuliani2024}. The framework is used to obtain best-effort near-term extrapolations by combining theoretical predictions with the latest experimental data. The individual model predictions are first preprocessed in a model orthogonalization step that mitigates the impact of model similarity before being mixed. This results in the only tunable hyperparameter of the method: the number of principal components used in the mixing stage. The performance of the BMC multi-model on the holdout validation set is then used to determine the number of components to avoid overfitting.
For the present study, the number of principal components was determined to be four.
Performance of the BMC 
multi-model prediction compared to individual models' predictions is shown in Fig. S3 of Supplementary Material~\cite{Sup}.

\begin{figure}
    \includegraphics[width=\linewidth]{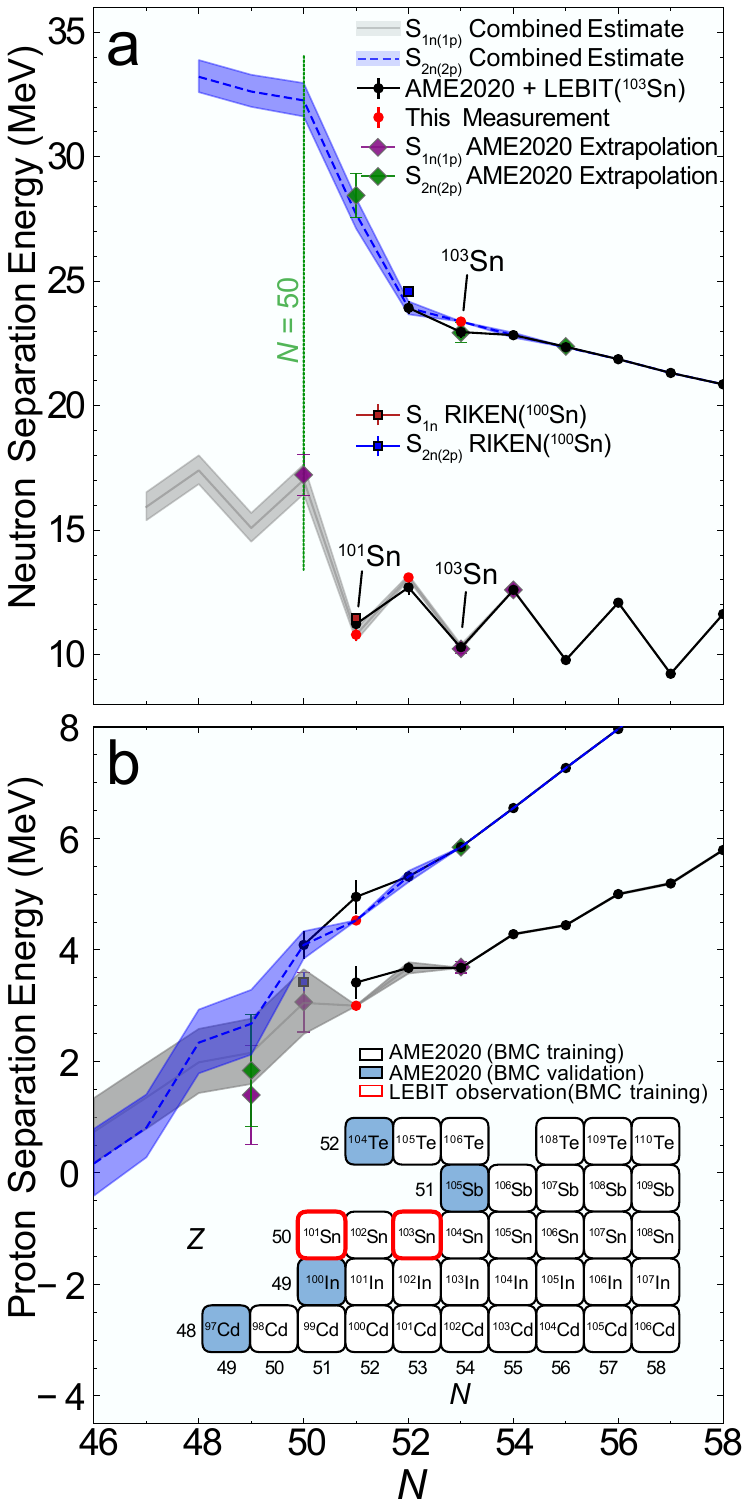}
    \caption{BMC predictions of separation energies when including recent experimental (training) data from LEBIT. Panels \textbf{a} and \textbf{b} display one (solid gray) and two (dashed blue) neutron and proton separation energies, respectively. Red dots correspond to the experimental values obtained in this work, while black dots represent AME2020 data~\cite{AME2020}, with the exception of ${}^{103}$Sn which is derived from Ref.~\cite{Ireland2025}. Diamonds represent AME2020 extrapolations for $S_{1n(1p)}$ (magenta) and $S_{2n(2p)}$ (green). Squares label $S_{1n}$ (brown) and $S_{2n(2p)}$ values obtained using ${}^{100}$Sn mass from Ref.~\cite{Lubos2019}. The bands mark the $1\sigma$ uncertainty from the statistical mean of the BMC prediction. The insert shows the set of nuclei used in BMC for training (white) and validation (blue), together with the training data points for ${}^{101,103}$Sn from this work and  Ref.~\cite{Ireland2025} (red outline). The vertical dashed green line indicates the $N = 50$ shell closure.}
    \label{fig:Sn_Sp_with_stats}
\end{figure}

For the training data, we used experimental binding energies reported in AME2020 for nuclei with $48\le Z \le52$ and $50\le N \le58$. The exceptions are the mass of $^{100}$Sn, which was excluded due to the ongoing discrepency in the reported $\beta$-delayed $Q$-values~\cite{Lubos2019, Hinke2012}, and the masses of $^{101,103}$Sn, which were taken from LEBIT mass measurements. The masses of the validation nuclei, $^{97}$Cd, ${}^{100}$In, ${}^{105}$Sb and ${}^{104}$Te, are taken from AME2020. The BMC validation set was chosen as the lightest experimentally measured nuclei in each respective isotopic chain and was not utilized either during training process or to refine the model predictions. Figure~\ref{fig:Sn_Sp_with_stats} shows the BMC mass extrapolations for $Z=50$ down to the proton drip line. The predictions are well-constrained outside the region $50\le N \le60$. Toward the $Z=50$ proton drip line, BMC predictions for the binding energy of $^{99}$Sn lie within 1.3~MeV (0.16\%) of the AME2020 systematic trends. We predict that the ${}^{96-99}$Sn isotopes are bound with respect to the emission of one proton, given that $S_{1p}>0$ is within the $1\sigma$ credible interval. Our calculations suggest that ${}^{96}$Sn is a two-proton drip line nucleus, likely with a fairly long lifetime (cf.~discussion in Ref.~\cite{Neufcourt2020}). Although the mean value $S_{2p}$ is slightly positive, the BMC prediction cannot exclude $S_{2p}<0$ within the $1\sigma$ credible
interval. With this in mind, we predict that the main decay mechanism for $^{96-99}$Sn is likely $\beta^+$-decay and electron capture, which is consistent with the previous multi-model study \cite{Neufcourt2020} employing the Bayesian Model Averaging method, which predicted the two-proton drip line for the $Z = 50$ chain to be at ${}^{96}$Sn. However, the BMA method's implicit assumption that ``truth’’ is contained within the model set is not typically valid in a nuclear physics context. Instead, combining models via BMC such as in this work outperforms BMA in both prediction accuracy and uncertainty quantification (see discussions in Refs.~\cite{Kejzlar2023, Giuliani2024}). Table~\ref{tab:BMC_predictions} summarizes the BMC energy extrapolations. 

\begin{table}[htbp]
    \centering
    \caption{Predictions for \(^{96-99}\)Sn binding energies $BE$ (in MeV), mass excess $ME$ (in keV), and separation energies (in MeV) using BMC. The recent LEBIT values for $^{101,103}$Sn were included in the training data.}
    \begin{tabular}{lrrrr}
        \hline
        \hline
        \textbf{Obs.} & \textbf{\(^{96}\)Sn} & \textbf{\(^{97}\)Sn} & \textbf{\(^{98}\)Sn} & \textbf{\(^{99}\)Sn} \\
        \hline
        $BE$  & 761.11(77) & 777.02(78) & 794.16(78) & 809.26(78) \\
        $ME$  & $-$25\,380(770) & $-$33\,220(780) & $-$42\,280(780) & $-$49\,320(780) \\
        S$_{1n}$  & - & 15.94(56) & 17.40(58) & 15.09(58) \\
        S$_{2n}$  & - & - & 33.22(65) & 32.62(64) \\
        S$_{1p}$  & 0.74(58) & 1.35(59) & 1.98(57) & 2.16(59) \\
        S$_{2p}$  & 0.17(60) & 0.81(56) & 2.34(57) & 2.68(58) \\
        \hline
        \hline
    \end{tabular}
    \label{tab:BMC_predictions}
\end{table}

To attempt to shed light on the $^{100}$Sn mass discrepancy reported in Ref.~\cite{Mougeot2021}, the BMC predictions for $\Delta_{3n}$ are plotted in Fig.~\ref{fig:3pointLEBIT}. BMC training and validation data are the same as in Fig.~\ref{fig:Sn_Sp_with_stats}.  Individual BMC frameworks were utilized for different observables given that they can have their own domains and measurement uncertainties depending on how they were determined. These BMC calculations predict a mass excess of $-58\,090(800)$~keV for $^{100}$Sn. This result agrees within the 1$\sigma$ credible interval with the mass excess of $-57\,150(240)$~keV reported in the AME2020, derived from the $\beta$-decay $Q$-value of $^{100}$Sn reported at GSI~\cite{Hinke2012}. On the other hand, the mass excess value for $^{100}$Sn of $-56\,490(160)$~keV derived from the more recent $\beta$-decay $Q$-value measurement at RIKEN~\cite{Lubos2019} lies outside of the $1\sigma$ BMC credible interval, albeit with a Bayes factor of $K\approx2$, hinting at only a minor preference for the measurement at GSI~\cite{kass1995bayes}. To test further, Fig.~\ref{fig:Sn_Sp_with_stats} also shows $S_{1n}$ and $S_{2n(2p)}$ separation energies with the ${}^{100}$Sn mass taken from the RIKEN study~\cite{Lubos2019}. Again, our results show a preference for the GSI value~\cite{Hinke2012} for ${}^{100}$Sn but can not conclusively exclude the results of Ref.~\cite{Lubos2019} from RIKEN. Our findings are in line with the recent \textit{ab-initio} calculations from Ref.~\cite{Mougeot2021}.

The stability of the BMC extrapolations can be probed by introducing or removing mass data in the training and calculating the changes to Table~\ref{tab:BMC_predictions} and the predicted mass of $^{100}$Sn. The predictions for $^{96-100}$Sn were thus repeated by exchanging the LEBIT mass data for $^{101,103}$Sn with the mass data reported in the AME2020. The individual changes to Table~\ref{tab:BMC_predictions} and the predicted mass value for $^{100}$Sn are $<0.1$\,MeV, demonstrating that the framework does not rely solely on systematic trends from neighboring nuclei.

An ambiguity still exists regarding the ground state spin and parity of ${}^{101}$Sn~\cite{seweryniak2007single, Darby2010, yakhelef2024neutron, PhysRevC.102.014304, fqfz-rf2c} that cannot be resolved with a precise mass value for $^{101}$Sn. Different theoretical models predict varying configurations in ${}^{101}$Sn. Shell-model calculations based on realistic interactions favor $7/2^+$ as the ground state of $^{101}$Sn and $5/2^+$ as the ground state of $^{103}$Sn~\cite{Darby2010}. \textit{Ab-initio} methods~\cite{PhysRevLett.120.152503} predict near degeneracy of both $2d_{5/2}$ and $1g_{7/2}$ states. With regards to our DFT predictions, both $2d_{5/2}$ and  $1g_{7/2}$ one-quasineutron states are close in energy. It should be noted, however,  that many-body correlations, such as pairing, are expected to significantly contribute to the ground state energy of ${}^{101}$Sn\cite{Darby2010}. To fully address this important question, a measurement of the magnetic moment of $^{101}$Sn is called for.

\textit{Conclusions}--In this work, the first high precision mass measurement of $^{101}$Sn is reported. This result represents an improvement over the prior precision by a factor of 300 and makes all nuclei in the \textsuperscript{109}Xe$\rightarrow$\textsuperscript{105}Te$\rightarrow$\textsuperscript{101}Sn $\alpha$-decay chain less bound than previously determined. The obtained lower mass uncertainty in $^{101}$Sn allows the discrepancy in the mass of $^{100}$Sn highlighted in Ref.~\cite{Mougeot2021} to be clearly visible in the three-point estimator for the neutron odd-even staggering at $Z=50$. The theoretical analysis performed around $^{101}$Sn represents the first application of the recently-developed BMC framework to real-world mass data of rare-isotopes. Our findings maintain support for \textsuperscript{96}Sn being a two-proton drip line nucleus, and show a slight preference for the mass value for \textsuperscript{100}Sn reported in the 2020 Atomic Mass Evaluation. This value was derived from the $\beta$-decay $Q$-value from GSI~\cite{Hinke2012} and is also supported by the theoretical study conducted in Ref.~\cite{Mougeot2021}. The results in this work demonstrate the importance of medium- and long range extrapolations of nuclear properties and the role that precise measurements, paired with statistical machine learning, can play in our confidence in those extrapolations. As experimental campaigns at rare-isotope facilities push closer to the particle drip lines, precise mass measurements will provide crucial observables for benchmarking the stability of state-of-the-art theoretical models. Among these, the  magnetic moment of $^{101}$Sn and the precise mass of \textsuperscript{100}Sn are long-awaited and will play a particularly pivotal role. 

\textit{Acknowledgments}--We thank Pablo Giuliani for useful discussions on the interpretation of errors in the BMC framework. We wish to also thank the FRIB operations team and acknowledge the specialized operational support provided by Sierra Rogers and Lily Stackable to make this measurement possible. This material is based upon work supported by the U.S. Department of Energy, Office of Science, Office of Nuclear Physics and used resources of the Facility for Rare Isotope Beams (FRIB) Operations, which is a DOE Office of Science User Facility under Award Number DE-SC0023633. This work was conducted with the support of Michigan State University, the US National Science Foundation under contracts no. PHY-2111185, the DOE, Office of Nuclear Physics under contract no. DE-AC02-06CH11357, DE-AC02-05CH11231, and DE-SC0022538. It was also supported by the U.S. Department of Energy, Office of Science and Office of Nuclear Physics under Awards Nos. DE-SC0023688  and DOE-DE-SC0013365, and DE-SC0023175 (Office of Advanced Scientific Computing Research and Office of Nuclear Physics, Scientific Discovery through Advanced Computing), DOE-DE-NA0004074 (NNSA, the Stewardship Science Academic Alliances program), and by the National Science Foundation under award number 2004601 (CSSI program, BAND collaboration). S.E.C. acknowledges support from the DOE NNSA SSGF under DE-NA0003960. W.S.P., A.M.H. and C.Q. acknowledge support from the U.S. National Science Foundation under grant number PHY-2310059. C.M.I. acknowledges support from the ASET Traineeship under the DOE award no. DE-SC0018362.

\section{Supplemental Material}--This text contains details relevant to the reported average mass ratio of $^{101}$Sn$^{2+}$ relative to $^{50}$Ti, the performance of individual energy density functionals on various mass filters, as well as the comparison of these models with that of the BMC multi-model.

\subsection{Individual mass measurement results}
\begin{figure}[h!]
\includegraphics[width=1\columnwidth]{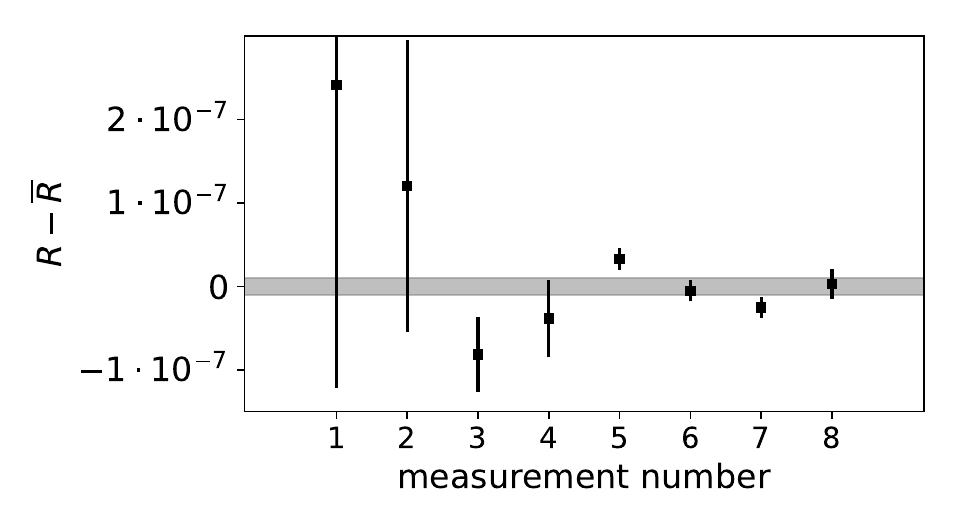}
\caption{The difference of frequency ratios \bm{$R$} and their weighted average $\bar R$ for the 8 measurements of \textsuperscript{101}Sn using PI-ICR. From left to right, two initial measurements of $\nu_{+}$ with $t_{acc_{+}}=$5.5 and 13.8~ms, two measurements with 50~ms, one with 150~ms, and three measurements with 163~ms of $\nu_{+}$ accumulation time were performed. The $\pm 1 \sigma$ error in the weighted average $\bar{R}$ is displayed by the gray band.}
\label{fig:101Sn_Ratios}
\end{figure}

The experimental systematic studies detailed in the main text determined the required shifts to the individually measured cyclotron frequency ratios $R$ of \textsuperscript{101}Sn\textsuperscript{2+} relative to \textsuperscript{50}Ti\textsuperscript{+}. Fig.~\ref{fig:101Sn_Ratios} compares these measurement results to the weighted average ratio $\bar{R}$ reported in the main text (data available upon reasonable request).

\subsection{Mass filters}
During the theoretical analysis, various mass indicators were employed. The one-nucleon separation energies in terms of mass excess $M_E$ are given by:
\begin{align}
\begin{aligned}
		S_{1p} (N, Z) &= ME(N,Z-1) - ME(N,Z) + m_{\rm H}, \\
        S_{1n} (N, Z) &= ME(N-1,Z) - ME(N,Z) + m_{\rm n},
		\label{eqs:separation_en}
  \end{aligned}
\end{align} 
where $m_{\rm H}$ and $m_{\rm n}$ are the hydrogen atom and neutron mass excess, respectively. The two-nucleon separation energies $S_{2p}$ and $S_{2n}$ are defined analogously. Additionally, we utilize the three-point estimator, $\Delta_{3n}(N,Z)= 0.5(-1)^N [ME(N-1,Z) - 2ME(N,Z) + ME(N+1,Z)]$, for the neutron odd-even staggering, selected due to its sensitivity to shell properties~\cite{Buskirk2024}. 

\begin{figure}
    \centering
    \includegraphics[width=1\linewidth]{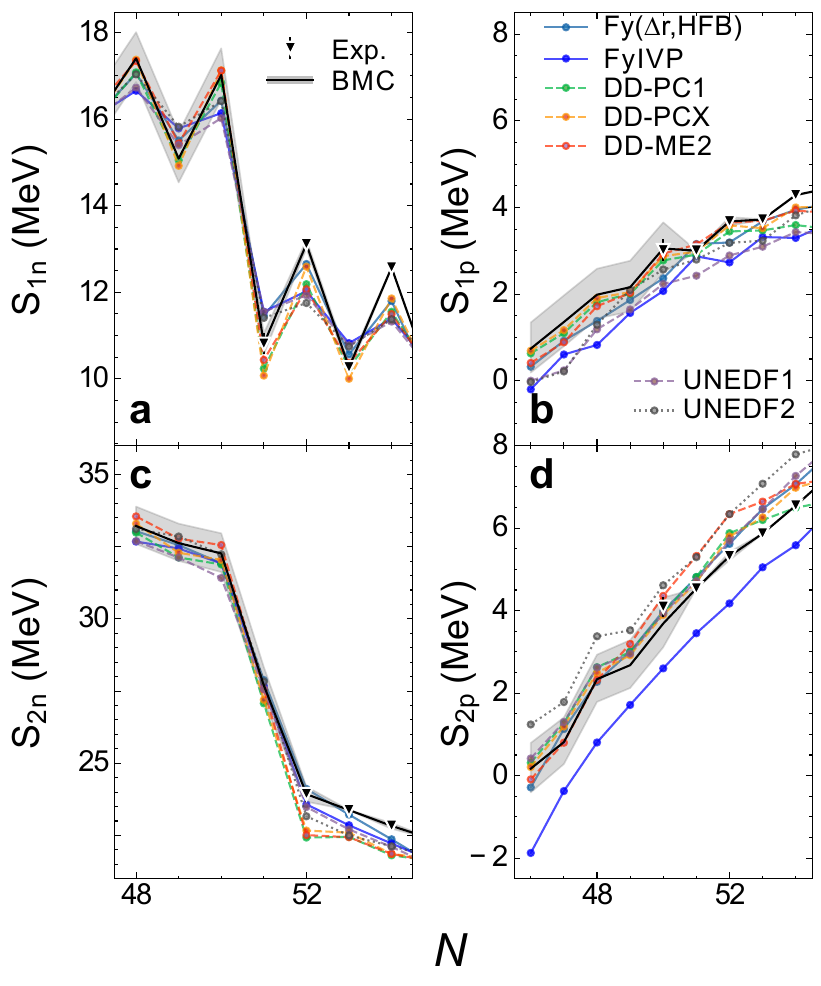}
    \caption{Tin separation energies of (a) one neutron, (b) one proton, (c) two neutrons, and (d)  two protons  calculated using the seven different EDFs utilized in this work, together with the BMC prediction and its $1\sigma$ uncertainty, compared to the experimental data taken from AME2020, this work, and Ref.~\cite{Ireland2025}.}
    \label{fig:separation_energies}
\end{figure}

The theoretical analysis was carried out using nuclear density functional theory (DFT) with state-of-the-art energy density functionals (EDFs). Figure~\ref{fig:separation_energies} shows one- and two-nucleon separation energies in the tin isotopic chain calculated using the seven different EDFs . Considering the calculations are performed in the neighborhood of a self-conjugated $N = Z$ nucleus, ${}^{100}$Sn, additional binding is provided by the Wigner energy term, added to the theoretical binding energies using the prescription of Refs.~\cite{Buskirk2024,Hamaker2021a}. The calculations predict small quadrupole deformations and soft potential energy surfaces for odd-$A$ tin isotopes. For instance, ${}^{101}$Sn has quadrupole deformation $\beta_2\sim 0.03$ consistently predicted with all models. In general, neutron-deficient tin isotopes exhibit fluctuations around the minimum energy due to collective ground-state correlations (GSC) \cite{Reinhard1987,Reinhard1999}. Based on the calculations for even-even tin isotopes with the Fy($\Delta r$,HFB) we estimate that GSC have a noticeable impact for $A > 106$ tin isotopes, being at most a 3 MeV correction to the total binding energy. Overall, the predictions between different models for proton and neutron separation energies of tin isotopes are fairly similar and in a good agreement with the main trends in the experimental data. In particular, for neutron separation energies, agreement between the Bayesian Model Combination (BMC) average and individual model predictions is excellent. On the other hand, somewhat larger discrepancy is obtained for one- and two-proton separation energies, especially for Fy(IVP).

\subsection{Bayesian model combination framework}
\begin{figure*}
        \centering
        \includegraphics[width=0.8\linewidth]{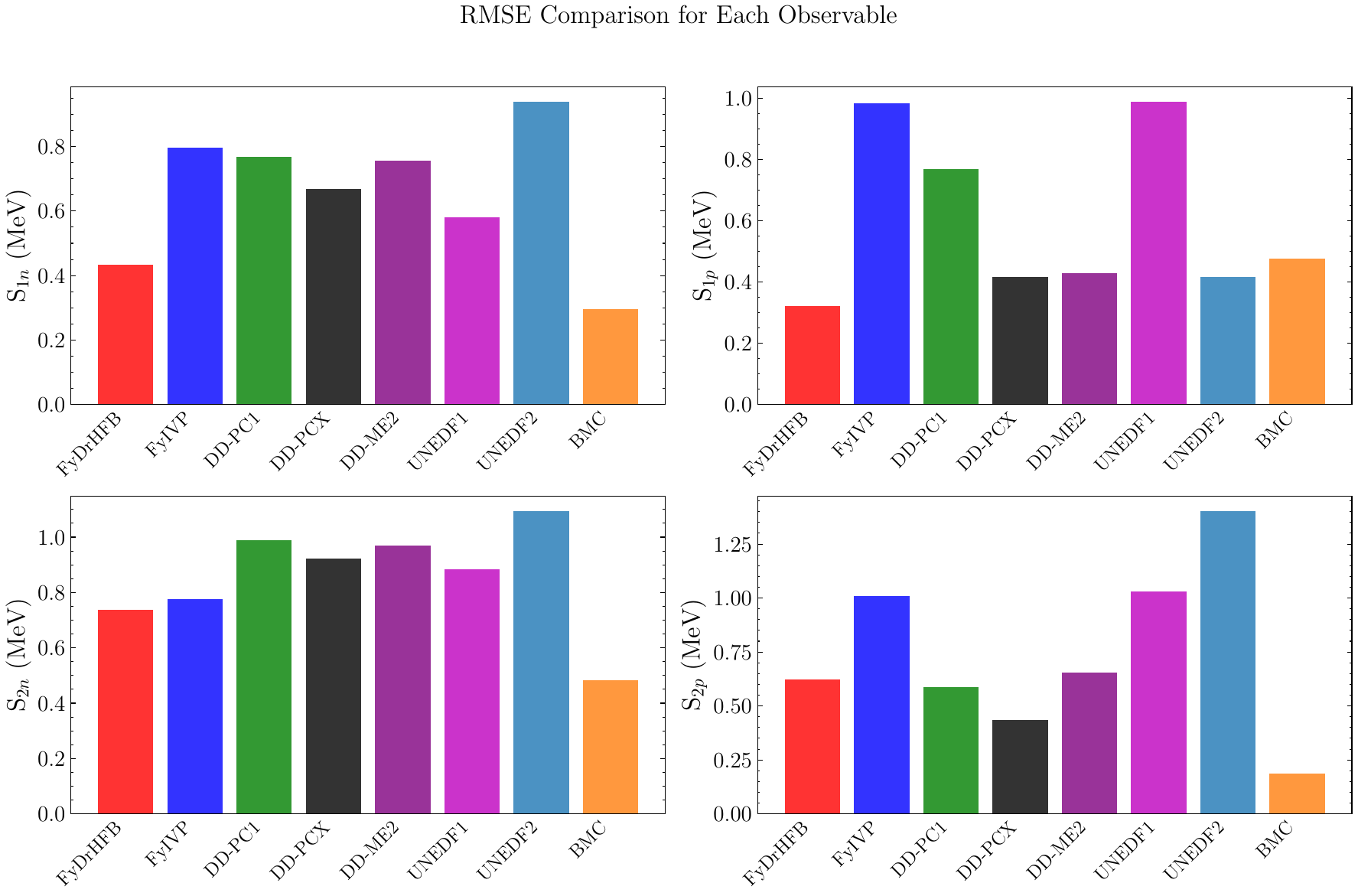}
        \caption{RMSE values for individual models as well as the BMC for the tin chain.}
        \label{fig:rmse_comparison}
\end{figure*}

The superiority of a principled BMC framework relative to its constituent computational models has been discussed in detail in Ref.~\cite{Giuliani2024}. A comparison of root-mean-square (RMS) deviations for the seven models utilized in this work as well as the BMC multi-model in the studied regions $52 \leq N \leq 58$ for the tin isotopic chain are shown in Figure~\ref{fig:rmse_comparison}. If the entire BMC training set used in this work and discussed in the main text were considered, then the resulting deviation would be lower for the BMC framework, as described in Ref.~\cite{Giuliani2024}. Even for a subset of the training set, however, BMC consistently performs better than the individual models, apart from $S_{1p}$ where it is similar to the best performing model. 

Additionally, as mentioned in the main text, the BMC validation set was chosen to contain the lightest experimentally measured nuclei in each respective isotopic chain. Although the mass of $^{100}$Sn is reported in the AME2020~\cite{AME2020} as derived from the $\beta$-decay value of Ref.~\cite{Hinke2012}, $^{100}$Sn is not included in the BMC validation set due to the ongoing discrepancy in its mass value. However, it should be noted that the validation nuclei are only used to tune the number of principal components kept in the model orthogonalization stage to avoid overfitting. Thus, the conclusions of this work are not influenced by either the inclusion or exclusion of any derived mass for $^{100}$Sn in the validation set.  

\providecommand{\noopsort}[1]{}\providecommand{\singleletter}[1]{#1}%

\end{document}